# Direct Observation of $d$-Wave Superconducting Gap Symmetry in Pressurized La$_3$Ni$_2$O$_{7-\delta}$ Single Crystals

Zi-Yu Cao[1,2,*], Di Peng[3,4,†], Seokmin Choi[2,*], Fujun Lan[4,*], Lan Yu[4,*], Enkang Zhang[5,*], Zhenfang Xing[4], Yuxin Liu[4], Feiyang Zhang[4], Tao Luo[4], Lixing Chen[5], Vuong Thi Anh Hong[2], Seung-Yeop Paek[2], Harim Jang[6], Jinghong Xie[1], Huayu Liu[1], Hongbo Lou[3,4], Zhidan Zeng[4], Yang Ding[4], Jun Zhao[5,†], Cailong Liu[1,†], Tuson Park[2,†], Qiaoshi Zeng[3,4,†], and Ho-kwang Mao[3,4]

[1] Key Laboratory of Quantum Materials Under Extreme Conditions in Shandong Province, School of Physics Science & Information Technology, Liaocheng University, Liaocheng 252059, China

[2] Center for Quantum Materials and Superconductivity (CQMS) and Department of Physics, Sungkyunkwan University, Suwon 16419, Republic of Korea

[3] Shanghai Key Laboratory of Material Frontiers Research in Extreme Environments (MFree), Shanghai Advanced Research in Physical Sciences (SHARPS), Shanghai 201203, China

[4] Center for High Pressure Science and Technology Advanced Research, Shanghai, China

[5] State Key Laboratory of Surface Physics and Department of Physics, Fudan University, Shanghai, China

[6] Institute of Nanostructure and Solid State Physics, University of Hamburg, Jungiusstrasse 11, 20355 Hamburg, Germany

*These authors contributed equally to this work.

[#]Correspondence should be addressed to di.peng@hpstar.ac.cn, zhaoj@fudan.edu.cn, cailong_liu@jlu.edu.cn, tp8701@skku.edu, or zengqs@hpstar.ac.cn.

**The recent discovery of superconductivity in pressure-stabilized bulk $La_3Ni_2O_{7-\delta}$, with a critical temperature ($T_c$) exceeding 77 K, has opened a new frontier in high-temperature superconductivity research beyond cuprates[1-8]. Yet, the superconducting gap amplitude and symmetry, the key parameters to characterize a superconductor[9-12], remain elusive due to the overwhelming challenges of gap studies under high pressure. Here, we introduce in situ directional point-contact spectroscopy conducted under truly hydrostatic pressure, enabling the direct mapping of the superconducting gap in pressurized $La_3Ni_2O_{7-\delta}$ single crystals. Depending on the junction orientation, differential conductance (d$I$/d$V$) spectra exhibit distinct V-shaped quasiparticle features and a sharp zero-bias peak, indicating a predominant $d$-wave-like pairing symmetry. Measurement of the $c$-axis gap amplitude ($\Delta$) yields a gap-to-$T_c$ ratio of $2\Delta/k_BT_c = 4.2(5)$, positioning $La_3Ni_2O_{7-\delta}$ firmly among unconventional, nodal high-$T_c$ superconductors. These findings set stringent constraints on theoretical models for nickelate superconductors and establish a robust spectroscopic approach for understanding superconductors under extreme pressures.**

The superconducting gap is the fingerprint of a superconducting state. In conventional superconductors, the superconducting gap exhibits an isotropic $s$-wave symmetry with a uniform phase across the Fermi surface. By contrast, high-temperature superconductors often exhibit anisotropic features with amplitude and phase variations, a hallmark of unconventional pairing driven by strong electronic correlations[9,10]. Among these systems, perovskite cuprates have remained a persistent focus of research, not only for the practical importance of their high $T_c$ but also for the theoretical interest of their well-documented unconventional $d_{x^2-y^2}$ gap symmetry[11,13]. Motivated by the cuprates, recent work on pressurized bulk $La_3Ni_2O_{7-\delta}$ with an analogous perovskite structure, showing high $T_c$ above liquid nitrogen temperatures and hints of unconventional pairing, has sparked intensive experimental and theoretical efforts to address their underlying superconducting pairing mechanism[1-7,14-26].

So far, the theoretical understanding of the pairing mechanism in pressurized $La_3Ni_2O_{7-\delta}$ remains intensely debated, with competing proposals of $s_{\pm}$-wave[20-25,27-31] and $d$-wave[15,32-35] anisotropic pairing symmetries. Unfortunately, typical gap measurements based on vacuum electron probes, such as angle-resolved photoemission spectroscopy (ARPES) and scanning tunneling microscope (STM), have been impossible under pressure due to their incompatibility with high-pressure devices. The recently developed high-pressure tunneling spectroscopy technique enables the direct measurement of an isotropic $s$-wave superconducting gap under extreme pressure[36,37]. Complementary spectroscopic methods, including Raman scattering, infrared spectroscopy, and ultrafast optical pump-probe spectroscopy, have also provided valuable insights into gap-like features under high pressure[38-40]. However, none of these approaches can resolve the momentum-dependent electronic structure under pressure, preventing experimental detection of the anisotropic superconducting gap as a key feature predicted in pressurized nickelate superconductors[38-40].

Therefore, developing in situ high-pressure techniques with momentum resolution are urgently required for a direct and definitive experimental determination of the superconducting pairing symmetry under extreme high pressures. To this end, directional point-contact spectroscopy (PCS) based on electric transport measurements on single crystals[12,41-43] offers a promising approach. In PCS, when a current flows across a normal metal/superconductor junction, Andreev reflection[12] occurs, where incoming electrons from the normal metal side with energies (applied voltages) smaller than the superconducting gap $\Delta$ are reflected as holes, and Cooper pairs are transmitted into the superconductor, showing enhanced conductance. At applied voltages larger than $\Delta$, the conductance decreases quickly to that of the normal state, allowing an accurate determination of the magnitude of $\Delta$. Moreover, PCS could also exhibit directional sensitivity. The angular dependence of quasiparticle injection is governed by the contact barrier $Z$. At $Z=0$, quasiparticles are transmitted across the interface with a uniform angular probability into the superconductor. As $Z$ increases, the transmission becomes more directional, favoring paths perpendicular to the interface[44]. By establishing ballistic scattering points on different crystallographic orientations and strategically maintaining maximized $Z$ by adjusting contact status using pressure, momentum resolution could be incorporated into PCS techniques for in situ high-pressure studies using a diamond anvil cell (DAC) (see schematic, Fig. 1).

Previous in situ high-pressure PCS measurements have mainly been conducted on relatively large samples using silver paste drops (tens of microns in size) as counter electrodes, a so-called 'soft'-PCS approach, in piston-cylinder clamp cells, where pressures are usually limited (below ~2 GPa)[45-48]. Therefore, this method is unsuitable for superconducting $La_3Ni_2O_{7-\delta}$ single crystals, which require high pressure exceeding ~14 GPa[1]. To overcome this limitation, we replicate the ambient-pressure needle-anvil method[12,41-43] and develop a directional in situ high-pressure PCS technique in a DAC, which employs sharpened Au tips in contact with tiny (tens of microns in size) superconducting single crystals of different crystallographic orientations.

The key to ensuring successful directional PCS measurement is the capability to gently bring the normal metal tip gradually into contact with a $La_3Ni_2O_{7-\delta}$ single crystal, forming a typically nanoscale ballistic contact in a controllable way within the very limited space of a DAC sample chamber. This can be achieved only by using a very soft pressure-transmitting medium to prevent the metal tip from severely compressing the sample surface. Our previous studies[3,49-51] have established a highly reproducible in situ hydrostatic pressure resistivity measurement technique using condensed helium as the best hydrostatic pressure transmitting medium in a DAC[52,53], therefore, providing a solid foundation for achieving controllable point contacts while maintaining excellent single-crystal integrity and bulk superconductivity at high pressures. In addition, another primary advantage of employing a hydrostatic pressure-transmitting medium in this PCS method is the ability to maximize the contact barrier Z for achieving momentum resolution by avoiding too close contact under high pressures, an outcome not achievable under non-hydrostatic conditions (see Methods for more details).

Using our developed directional in situ high-pressure PCS technique in a DAC using helium as the hydrostatic pressure-transmitting medium, we explored the differential conductance ($dI/dV$) spectra of $La_3Ni_2O_{7-\delta}$ on high-quality single crystals synthesized via the high-pressure optical-image floating zone technique[3]. Although typical PCS measurements require only three electrical contacts, we adopted a four-point geometry on each $La_3Ni_2O_{7-\delta}$ single crystal, allowing for multiple independent PCS data acquisitions while enabling four-probe resistance measurements under the same experimental conditions to confirm the pressure-induced superconductivity (zero resistance), ensuring repeatability. Three independent pressure cells (pressure range: 20-22 GPa) were designed for the PCS studies, providing nine distinct measurement points (see Extended Data Fig. 1 for more experimental details). All samples exhibited relatively sharp superconducting transitions to zero-resistance states, with $T_c$ ranging from 64 K to 68 K, and a transition width of less than 10% of $T_c$ (Fig. 2), demonstrating the consistent high quality of the $La_3Ni_2O_{7-\delta}$ single crystals.

Figure 3 shows representative directional PCS spectra at 20-22 GPa and low temperatures (2 K or 3 K). The $dI/dV$ curves were obtained by numerical differentiation of $I$-$V$ characteristics and normalized by their median value for comparison. With point contact along the $c$-axis, we observe symmetrical peaks at ± 9.7 meV and a V-shaped dip crossing zero bias (Fig. 3a). With point contacts perpendicular to the $c$-axis, we reveal a pronounced zero-bias conductance peak with subtle gap edges (Fig.3c) in one orientation, and a combination of zero-bias conductance peak plus partially developed symmetric gap features (Fig.3b). The spectra in Fig. 3c are in reasonable agreement with the 2-dimensional Blonder-Tinkham-Klapwijk (2D-BTK) model for the $d$-wave nodal direction density of states[54,55]. The spectrum in Fig. 3c is similar to that observed in previous PCS studies on $Yba_2Cu_3O_7$ with current injection between the nodal and antinodal directions[12,55,56]. These directional dependences align well with the $d$-wave BTK model predictions illustrated in Fig. 1, showing distinguishing nodal and antinodal PCS signals observed in canonical *d*-wave superconductors, as obtained before using directional PCS or scanning tunneling microscopy (STM)[57-62].

Figure 4 presents the temperature- and magnetic field-dependent PCS results with ballistic contacts along the *c*-axis at 22.0 GPa. At 2 K, d*I*/d*V* exhibits symmetrical coherence peaks at ± 9.7 meV, which gradually merge to a diminishing single peak approaching $T_c$. No pseudogap features were detected above $T_c$. Considering that 9 T magnetic field exhibits limited suppression of superconductivity in the $La_3Ni_2O_{7-\delta}$ resistance at 2 K, we also measured PCS data with applied magnetic fields at 30 K to investigate the superconducting gap evolution with magnetic field. We have quantified the coherence peak separation as the superconducting energy gap 2Δ, and tracked its evolution with temperature and magnetic field (Figs. 4b, d). The data closely follow the BCS prediction (dashed curves), with 2Δ = 19.4 meV, yielding $2\Delta/k_BT_c$ = 3.72. Similar results were confirmed across *c*-axis contacts on Samples #1 and #3. Four contacts showed V-shaped conductance structures (Extended Data Figs. 2 and 3), and the averaged gap-to-$T_c$ ratio is 4.2(5). The reproducibility and consistency of the characteristic features in conductance spectra across different crystallographic orientations and among various samples demonstrate their intrinsic origin and spectroscopic reliability. We attribute the slightly lower values in Sample #2 to local $T_c$ suppression from relatively excessive uniaxial tip pressure (Extended Data Fig. 4).

Although theoretical models have proposed both *s*- and *d*-wave pairing symmetries as possible candidates for nickelate superconductors, our experimental findings impose stringent constraints. The clear manifestation of *d*-wave symmetry, combined with the measured coupling ratio $2\Delta/k_BT_c$ of 4.2(5), significantly narrows the parameter space for viable theoretical models. Thus, our results can serve as critical benchmarks for discriminating between the competing theoretical proposals[15,32-35].

The identification of the *d*-wave pairing symmetry in $La_3Ni_2O_{7-\delta}$ carries profound implications for the field of high-temperature superconductivity. First, it strongly reinforces the conceptual parallel between nickelates and cuprates despite their chemical and structural differences. Both systems host layered perovskite frameworks, exhibit strong electron correlations, and are situated near antiferromagnetic instabilities, together with a high $T_c$ exceeding 77 K. The explicit experimental evidence of a nodal superconducting gap in $La_3Ni_2O_{7-\delta}$ places it within the same symmetry class as the cuprates and suggests that the essential ingredients for *d*-wave pairing may be more broadly realized across different material platforms.

Moreover, the fact that $La_3Ni_2O_{7-\delta}$ exhibits a *d*-wave gap despite possessing a bilayer structure with strong interlayer hybridization[1,19,24,25] distinguishes it from quasi-two-dimensional cuprates with relatively weak interlayer coupling. This result implies that the pairing interaction in $La_3Ni_2O_{7-\delta}$ is sufficiently robust to withstand the band-splitting and $k_z$ dispersion introduced by interlayer coupling, expanding the theoretical frameworks in which *d*-wave superconductivity can arise. In this context, $La_3Ni_2O_{7-\delta}$ serves as a compelling testbed for examining how dimensionality, multiple-orbital character, and hybridization influence superconducting symmetry.

In summary, we report the first experimental determination of the superconducting gap symmetry of $La_3Ni_2O_{7-\delta}$ single crystals using in situ high-pressure directional PCS under extreme hydrostatic pressure. By creating ballistic contacts between Au tips and the $La_3Ni_2O_{7-\delta}$ sample strategically oriented parallel and perpendicular to the *c*-axis with a maximized contact barrier Z, we achieved controlled quasiparticle injections along a (mainly) specific *k*-space axis using condensed helium as the hydrostatic pressure-transmitting medium in a DAC. This method enables momentum-resolved spectroscopic determination of an anisotropic superconducting order parameter in pressurized $La_3Ni_2O_{7-\delta}$, revealing a predominantly *d*-wave pairing symmetry, with the average gap-to-$T_c$ ratio ($2\Delta/k_BT_c$) of 4.2(5). This advance not only provides critical insights into the elusive pairing mechanism in $La_3Ni_2O_{7-\delta}$ but also establishes a general in situ spectroscopic approach to probing strongly correlated high-$T_c$ superconductors under extreme conditions.

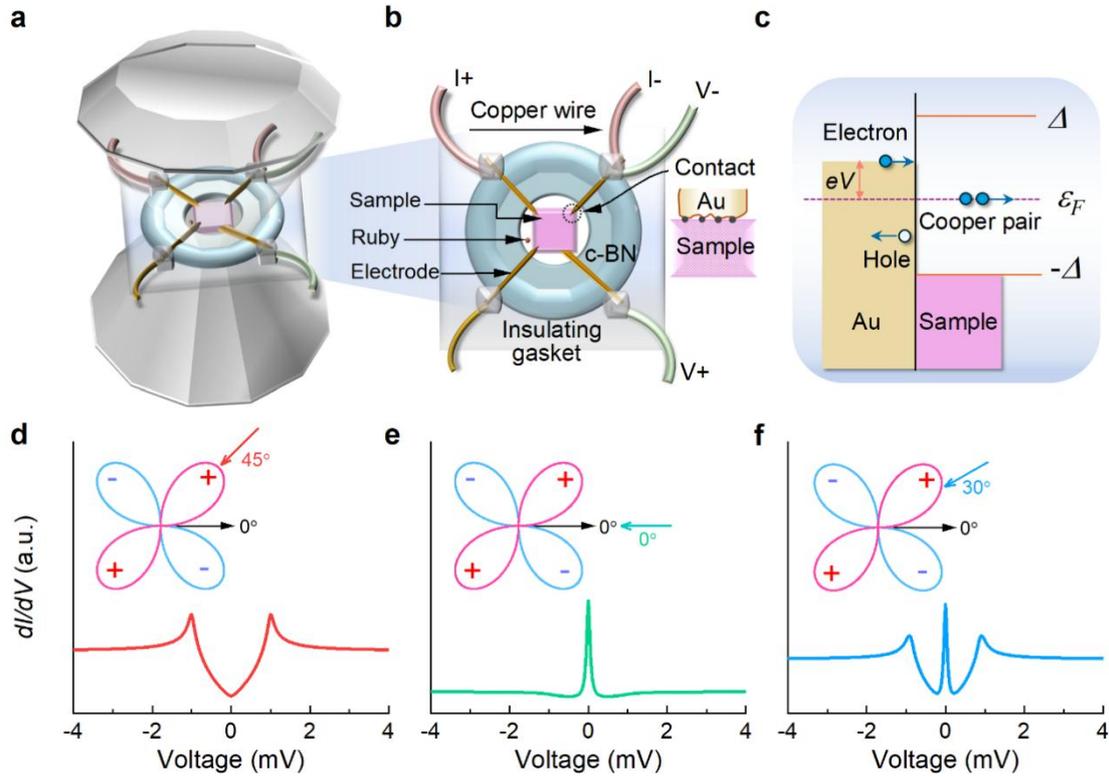

**Figure 1| Schematic diagram of directional point-contact spectroscopy in a diamond anvil cell. a,** Schematic diagram of the point-contact spectroscopy setup constructed within a diamond anvil cell. Triple Au probes established *ab*-plane contacts. Sample chamber shrinkage caused by pressure-triggered helium volume reduction forces another Au probe to slide toward the sample, with side contact achieved at target pressures. **b,** Top-view schematic of differential conductance (d$I$/d$V$) or resistance characterization inside the sample chamber. The right panel depicts a conceptual schematic of the realistic Au/La$_3$Ni$_2$O$_{7-\delta}$ point-contact junction, where multiple short-circuit points (black dots) contact the measured sample. **c,** Andreev reflection dynamics at Au/La$_3$Ni$_2$O$_{7-\delta}$ interface. **d-f,** 2D-BTK simulation results using barrier strength Z = 10 and gap magnitude Δ = 1 meV for a *d*-wave superconductor, showing quasiparticle incidence along the antinodal, nodal, and 30° off-nodal directions, respectively.

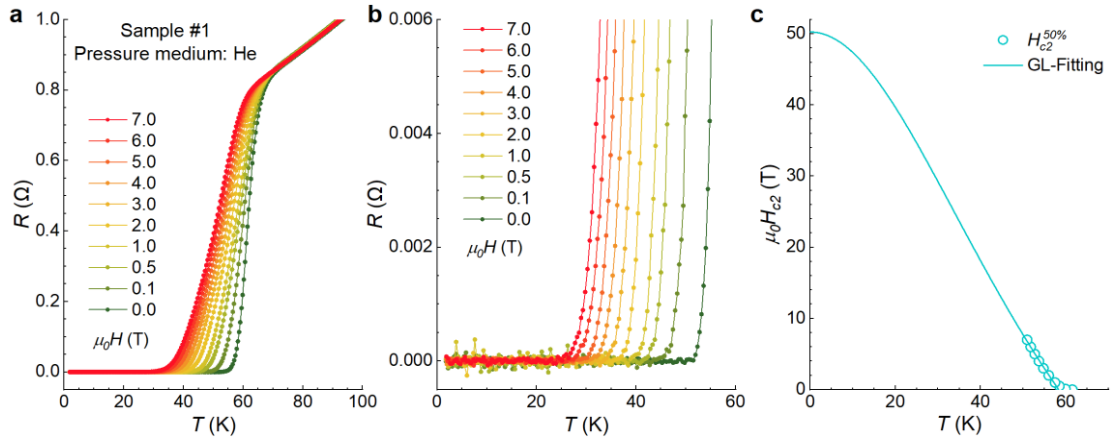

**Figure 2| Characterization of superconductivity in $La_3Ni_2O_{7-\delta}$ under pressure. a,** Temperature-dependent resistance measured in Sample #1 under various magnetic fields applied along the *c*-axis direction at 22.0 GPa. At zero field, the superconducting transition width is less than 10% of $T_c$ onset, marking the sharpest transition observed in nickelate superconductors to date. After the sharp drop, the resistance vanishes immediately, indicating a relatively high superconducting volume fraction in the measured sample. **b,** Zoomed view of the zero-resistance regime in **a**. **c,** The upper critical field, extracted from **a,** as functions of temperature. The dashed line represents the G-L fitting, giving an $H_{c2}(0)$ of 50.3 T. The criterion for $T_c$ is the 50% drop in resistance.

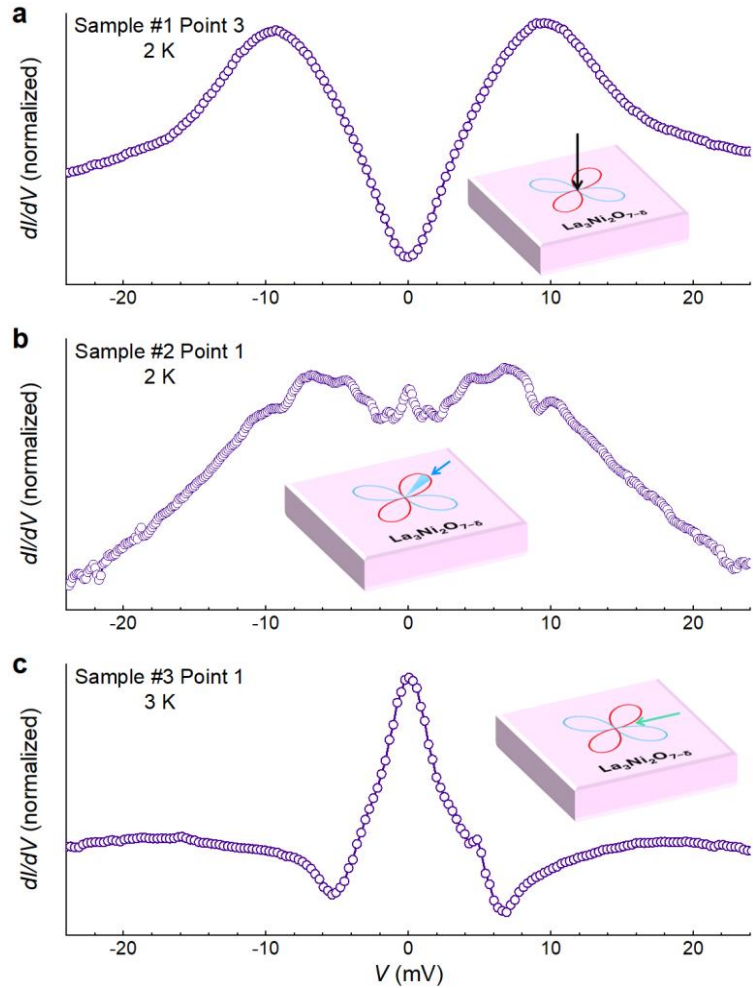

**Figure 3| Directional PCS on Au/La₃Ni₂O₇₋δ contacts at specified crystal orientations, measured at 2 K or 3 K and at 20-22 GPa. a,** The *dI/dV* measurement taken at contact point 3 with *c*-axis orientation of Sample #1 at 22.0 GPa. Based on Andreev reflection, the conductance of the junction turns out to be doubled for $V < \Delta/e$. However, the order parameter anisotropy and interfacial barrier Z jointly modify quasiparticle injection, manifesting as a characteristic V-shaped differential conductance spectrum in the *ab*-plane junction. **b,** The directional PCS measurements along the antinodal direction (visually aligned) of Sample #2 at 20.7 GPa. **c,** The directional PCS measurements along nodal direction of Sample #3 at 20.3 GPa. Within a *d*-wave framework, the zero-bias conductance peak is attributed to zero-energy Andreev bound states. The dip symmetrically appearing at the gap edge is distinct from that in an s-wave superconductor

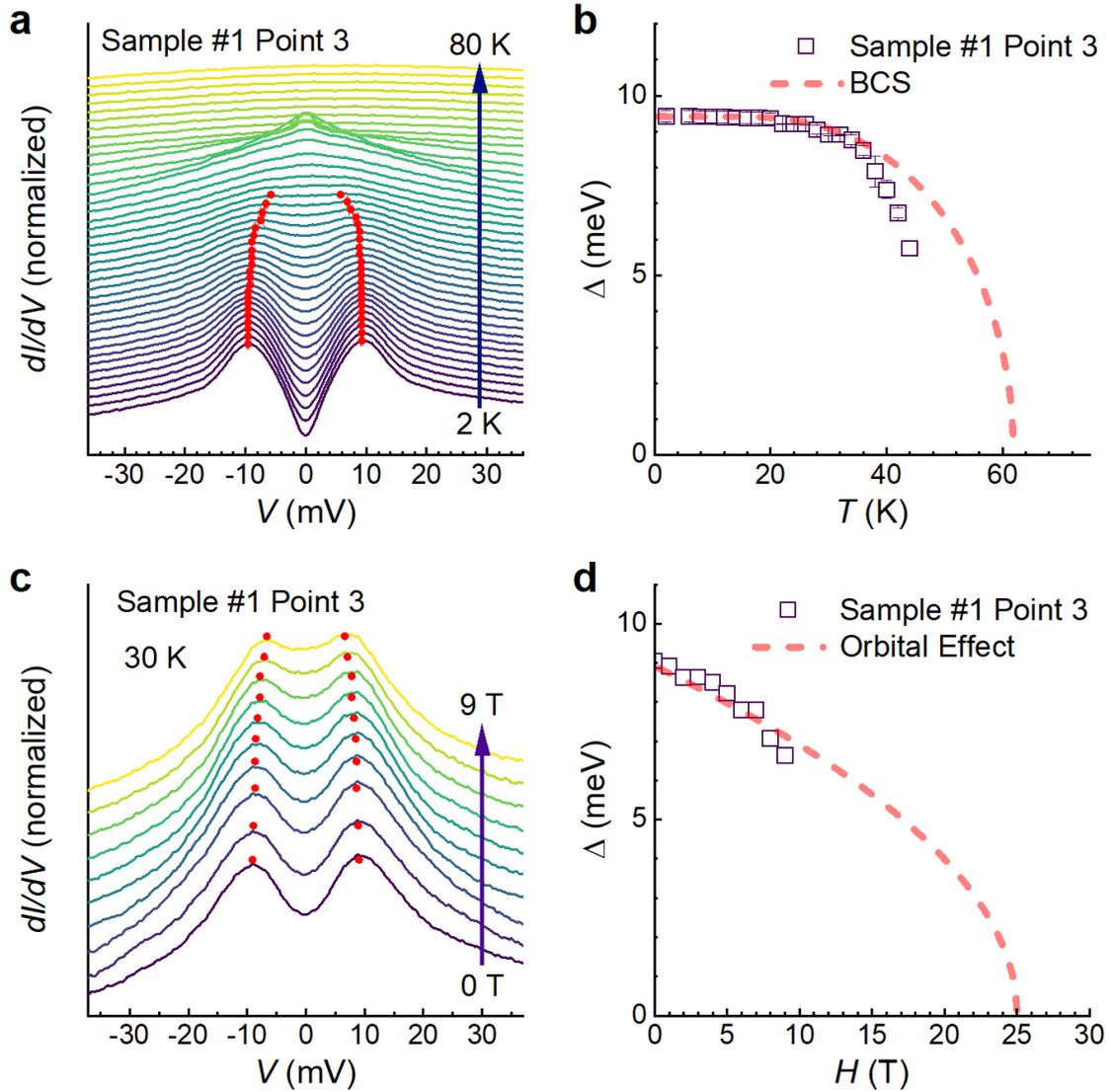

**Figure 4| Order parameter amplitude and its temperature and field evolution at 22.0 GPa. a,** Normalized d$I$/d$V$ measured at contact point 3 in Sample #1 as a function of temperature (2–80 K). Raw data are provided in the Supplementary Information (SI). Red dots indicate the coherence peak positions, where the peak-to-peak width yields the 2Δ. **b,** Temperature dependence of the Δ extracted from a. The dashed red line is the BCS fitting. **c,** Normalized d$I$/d$V$ spectra at point 3 in Sample #1 at 30 K under applied magnetic fields ranging from 0 to 9 T. Based on the $H_{c2}$ measurements in Fig. 1, the effect of a 9 T magnetic field on the superconducting gap at 2 K is negligible. We therefore performed measurements at 30 K, where the superconducting gap is more susceptible to a magnetic field. **d,** The suppression of the superconducting gap under a magnetic field. The dashed red line represents the estimated power-law behavior: $\Delta(H, 30\text{ K}) = \Delta(0, 30\text{ K}) \cdot (1 - H/H_{c2})^{1/2}$, where $H_{c2}$ is determined from resistance measurements.

# Methods

**Single-Crystal Growth of Samples.** The precursor powder for $La_3Ni_2O_7$ was synthesized using standard solid-state reaction techniques. $La_2O_3$ (Aladdin, 99.99%) and NiO (Aladdin, 99.99%) were used as raw materials, with an additional 0.5% NiO added to compensate for the potential volatilization during crystal growth. Before weighing, the $La_2O_3$ powder was dried to eliminate the moisture absorbed during storage. The mixture was then finely ground and mixed, followed by calcination at 1373 K for 24 h. This procedure was repeated twice to ensure a complete and homogeneous reaction. The powder was subsequently pressed into polycrystalline rods (~12 cm long and ~6 mm in diameter) under a hydrostatic pressure of 70 mPa for 30 min, followed by sintering at 1673 K in air for 12 h. Single crystals were cultivated in a vertical optical-image floating-zone furnace (Model HKZ, SciDre) at Fudan University. $La_3Ni_2O_7$ single crystals were grown under an oxygen pressure of 14-16 bar and a 5 kW xenon arc lamp. The growth rate was 3–5 mm/h after a rapid growth procedure at 20 mm/h to improve the density.

**Diamond anvil cell high-pressure technique.** Several standard-cut diamond anvils with ~400 μm culets were employed for the project. The experimental gaskets were fabricated by first laser-drilling 400 μm through-holes in pre-indented rhenium sheet, then packing the holes with the mixture insert of cubic boron nitride nanoparticles and epoxy (c-BN+epoxy) under high pressure, and subsequently laser-drilling 160 μm in diameter holes through the packed c-BN+epoxy layer. Helium was employed as the pressure-transmitting medium to optimally maintain hydrostatic conditions. The pressure was calibrated using the ruby fluorescence peak. A BeCu cell with an outer diameter of ~23 mm was employed for the low-temperature experiments.

**Resistance measurements under high pressure and low temperature.** A low-temperature environment was provided by a helium-4 closed-cycle refrigerator and a physical property measurement system (PPMS-9, Quantum Design Inc.). Electrical resistivity was measured using the van der Pauw lock-in technique at 17 Hz with a Lakeshore 370 AC resistance bridge. A current of 316 μA to 1 mA was applied for the resistivity measurements at low temperatures. The temperature was monitored using a resistance thermometer. The ramping rate was maintained at 0.1-0.5 K/min during cooling and warming procedures using a Lakeshore 350 cryogenic temperature controller.

**In situ high-pressure directional point-contact spectroscopy.** All ballistic contacts were fabricated manually. Specifically, we carefully optimized the initial thicknesses of the sample, gasket, and Au tips at ambient conditions, and then loaded the sample with four Au leads into a DAC. Ideally, the initial spatial gap between the Au tip and the sample should be set such that direct contact is avoided at low pressures but nearly gets closed at target pressures in a gradual way, a requirement that can be well achieved with He as the pressure-transmitting medium. During careful, incremental compression, the sample chamber thickness gradually shrinks between the two diamond anvils, diminishing the spatial gap between the Au tip and the sample to eventually form a ballistic contact between them, with a contact size smaller than the electron mean free path. Crucially, this could be achieved only by preserving the nanoscale surface roughness of the Au tip at target pressures. For each specific sample, too low or too high pressure could both potentially result in experimental failure due to contact loss during warming or excessive contact pressure-induced non-ballistic (diffusive or thermal) contact.

Moreover, to fabricate contact along the $c$-axis of a single crystal sample, the sample chamber deformation (shrinkage) caused by pressure-induced helium volume reduction forces the side-contact Au probe to gradually slide toward the sample lateral surface, with side contact achieved at target pressures. In this study, we employed three Au probes along the $c$-axis and a single Au contact perpendicular to the $c$-axis. The nodal and antinodal directions according to the single crystal orientations were measured using separate pressure cells.

The voltage responses to the direct current (DC) were measured at three small current steps ($\Delta I$ is less than 0.5% of the whole curve), and the slope was calculated to obtain the differential conductance ($dI/dV$) value. The corresponding bias voltage at the junction was estimated as the average of the three values. All $dI/dV$ values were obtained from the 3-point moving average to minimize the effects of electromotive force in the circuit. The contact resistance in all experiments ranges between 5 and 20 ohms.

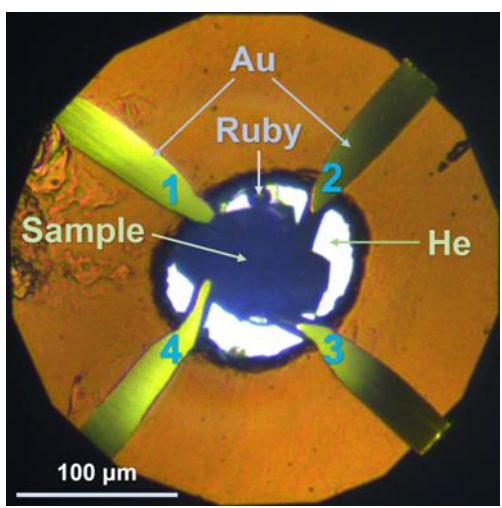

**Extended Data | Fig. 1. A photograph of a $La_3Ni_2O_7$ single crystal sample with four electrodes for in situ high-pressure resistance and directional PCS measurements loaded in a DAC.** The probes were labeled as probe 1–4. Probe 4 was first pressed hard against the sample using a diamond anvil to fix the sample's position in the sample chamber while the sample was being compressed in the He medium. Through the diamond window, probe 4 exhibits a shiny surface at its tip, indicating a relatively high axial pressure (similar to typical cases using solid pressure-transmitting media). In this case, the point contact in probe 4 may have numerous parallel contacts with relatively large contact sizes (sometimes larger than the electron mean free path) owing to the severe deformation of the soft Au tip by axial stress. Non-hydrostatic pressure may induce gap closure below $T_c$, as shown in Extended Data Fig. 4. Indeed, the PCS data collected by probe 4 were always in the thermal region, which did not reflect the nature of the intrinsic gap. Probes 1 and 2 make contacts along the *c*-axis (i.e., on the *ab* plane). These electrodes did not exhibit shiny surfaces in the contact area, indicating the absence of obvious axial stress from the diamond anvil to the Au probe. Probe 3 was placed in contact perpendicular to the *c*-axis. The sample chamber deformation caused by the pressure-induced reduction in helium volume forces probe 3 to slide toward the sample's lateral surface, with the side contacts touching different lattice orientations at the target pressures. Owing to the softness of Au, probes 1-3 have gentle contacts with the sample that do not empirically damage the Au nanoscale surface morphology, enabling reliable ballistic scattering spectroscopic data under pressure. Except for probe 4, each cell may ideally have three points available to conduct PCS measurements. However, the mechanical stability perpendicular to the *c*-axis contact (probe 3) is relatively weak during pressure and or temperature variations. Currently, we can only conduct PCS measurements for probe 3 at low temperatures.

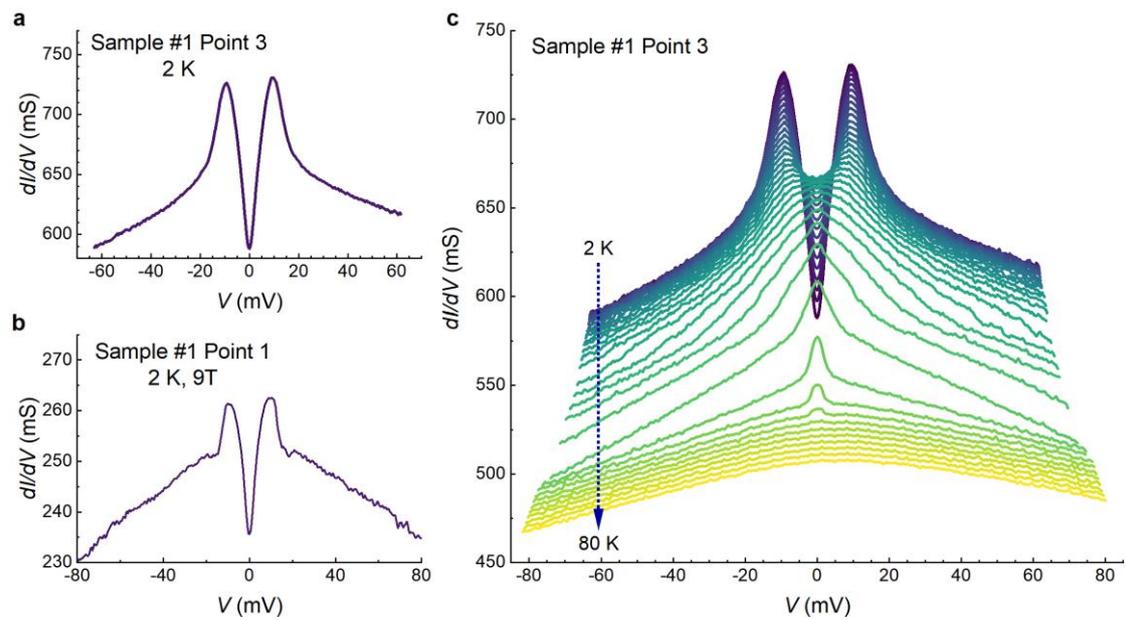

**Extended Data | Fig. 2. Along *c*-axis PCS measurement on sample #1 at 22.0 GPa. a,b,** Raw PCS data measured on sample #1, where point-1 and point-3 are independent contact points along the *c*-axis, exhibiting well-defined gap features. **c,** PCS measurements of sample #1 point-3 at temperatures ranging from 2 K to 80 K.

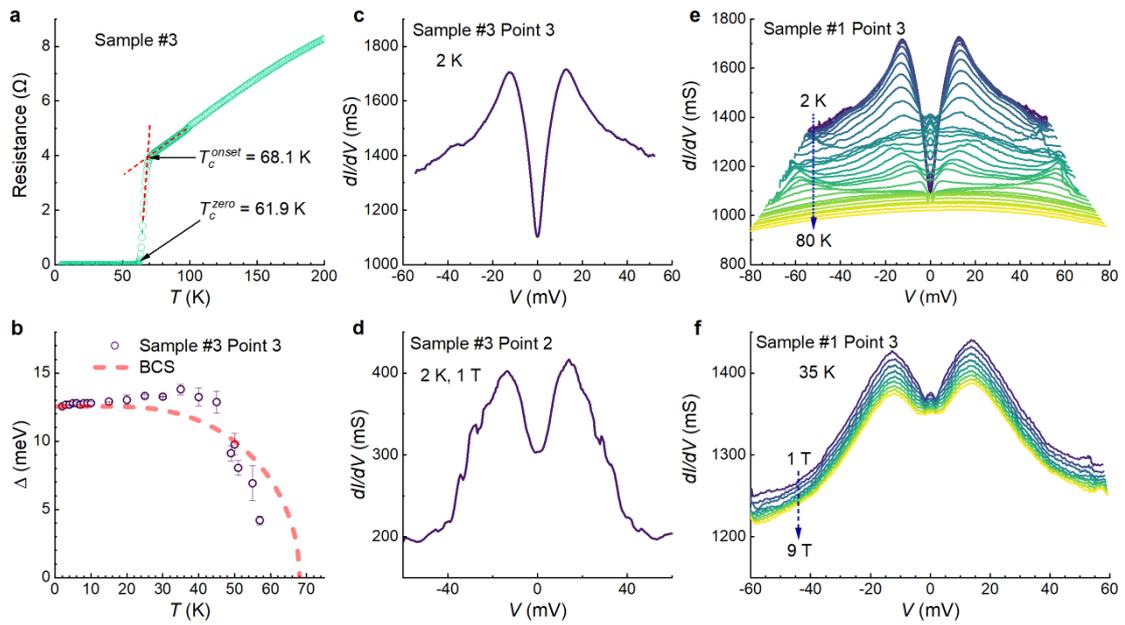

**Extended Data | Fig. 3. Along the *c*-axis PCS measurement on sample #3 at 20.3 GPa. a,** Temperature-dependent resistance of sample #3, showing $T_c$ onset at ~68.1 K and $T_c$ zero at ~61.9 K. **b,** Temperature evolution of superconducting gap derived from data in **e**. The dashed line corresponds to the BCS fit. **c,d,** Typical raw PCS data measured on sample #3. **e,f,** PCS measurements on sample #1 and point-3 at various temperatures and magnetic fields.

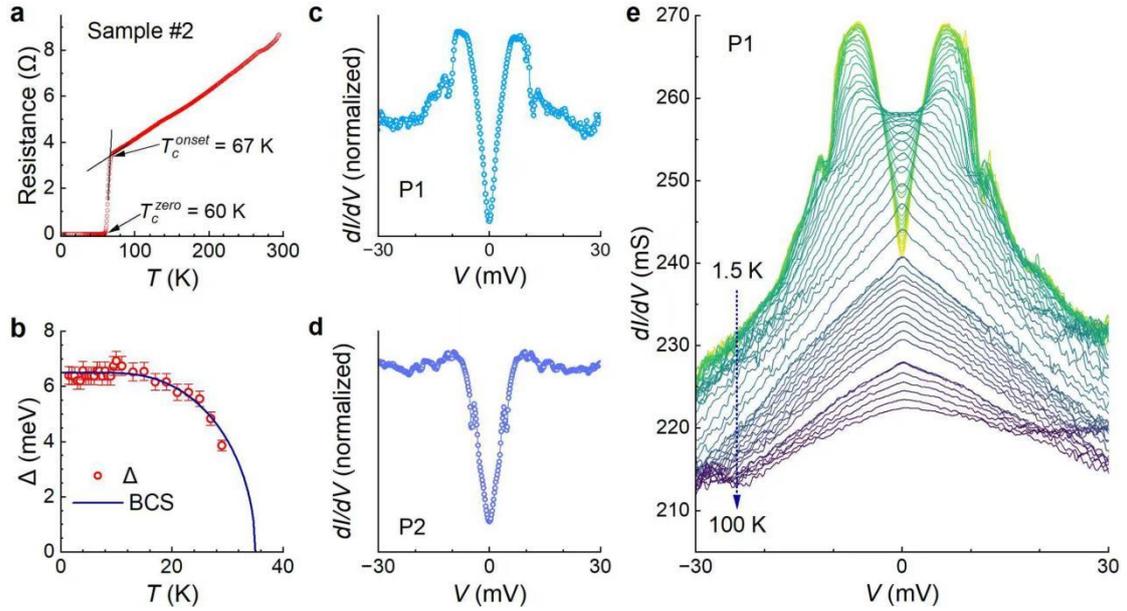

**Extended Data | Fig. 4. Along c-axis PCS measurement on sample #2 at 20.7 GPa. a,** Temperature-dependent resistance of sample #2. A sharp drop with $T_c$ onset at ~67 K is observed. **b,** Temperature evolution of the superconducting gap derived from data in **e**. Clearly, the extrapolated gap closure occurs at a temperature substantially below the $T_c$ determined in **a**. We attribute this occasional discrepancy to the development of non-hydrostatic conditions at the contact induced by excessive axial pressure from the Au tip, which may locally suppress $T_c$ at the contact point. In other experiments, special attention was paid to avoid such excessive axial compression from the Au tip. As a result, this issue of the energy gap closing below $T_c$ could be effectively resolved. **c,d,** Normalized $dI/dV$ measured at point-1 and point-2. **e,** The raw $dI/dV$ data of point-1 at various temperatures from 1.5 to 100 K.